\documentclass[a4paper,11pt]{article}
\pdfoutput=1
\usepackage{jcappub}

\usepackage{graphicx,epsfig}
\usepackage{subfig}
\usepackage{amsmath}
\usepackage{color}
\usepackage[normalem]{ulem}
\usepackage{natbib}
\usepackage{dcolumn}
\usepackage{amssymb}
\usepackage{amsfonts}
\usepackage{amsbsy}
\usepackage{rotating}
\usepackage[english]{babel}

\title{\boldmath BAO Cosmography}

\author[a]{Ruth Lazkoz,}
\author[b]{Jailson Alcaniz,}
\author[a]{Celia Escamilla-Rivera,}
\author[a]{Vincenzo Salzano}
\author[a]{and Irene Sendra}

\affiliation[a]{Fisika Teorikoaren eta Zientziaren Historia Saila, Zientzia eta Teknologia Fakultatea, \\ Euskal Herriko Unibertsitatea, 644 Posta Kutxatila, 48080 Bilbao, Spain}
\affiliation[b]{Observat\'orio Nacional, 20921-400, Rio de Janeiro - RJ, Brasil}

\abstract{Cosmography provides a model-independent way to map the expansion history of the Universe. In this paper we simulate a \textit{Euclid}-like survey and explore cosmographic constraints from future Baryonic Acoustic Oscillations (BAO) observations. We derive general expressions for the BAO transverse and radial modes and discuss the optimal order of the cosmographic expansion that provide reliable cosmological constraints. Through constraints on the deceleration and jerk parameters, we show that future BAO data have the potential to provide a model-independent check of the cosmic acceleration as well as a discrimination between the standard $\Lambda$CDM model and alternative mechanisms of cosmic acceleration.}


\begin{document}
\maketitle
\flushbottom

\section{Introduction}

Cosmic acceleration \cite{SNfirst,SNsecond} is certainly one of the most remarkable cosmological findings of the last century. It is well known, however, that all evidence we have so far for this ``dynamical phase transition" in the cosmic expansion are indirect in the sense that they appear only in the context of a given cosmological scenario.

In this regard, the determination of cosmographic parameters as well as their time evolution provide a unique method to map the cosmic expansion in a model-independent way and constitutes one of the major challenges in observational cosmology. Future surveys, like {\textit{Euclid}}~\cite{euclid} and J-PAS~\cite{Moles:2009db}, may provide a variety of high-precision observations, which includes distance measurements to Type Ia Supernovae, galaxy cluster abundance and measurements of the baryonic acoustic oscillations (BAO), a generic feature of the power spectrum of large-scale structure. In particular, BAO observations are valuable tracers of the cosmic expansion and a powerful method for setting cosmological constraints (see, e.g., \cite{Eisenstein,bao,wiggle} and references therein).

Our goal in this paper is to discuss the expected constraints of a cosmographic and, therefore, model-independent analysis of future BAO observations. To this end, we use Monte Carlo (MC) simulations of expected BAO data from a {\textit{Euclid}-like} survey~\cite{euclid}. Future BAO data from this kind of mission are supposed to be much more accurate than the present data used in cosmological analysis and must reach redshift $z \approx 2$, which is comparable to the limit achievable nowadays with Type Ia supernovae. We use synthetic BAO data to investigate the optimal order of the cosmographic expansion that must be used in order to reliably constrain the main cosmographic parameters and map the cosmic expansion.

\section{Cosmography}

Cosmography relies on the assumption that the spacetime geometry is well described on large scales by the Friedmann-Robertson-Walker line element. The key relation for this kind of approach is the Taylor expansion of the scale factor:
\begin{eqnarray}\label{eq:aseries}
\frac{a(t)}{a(t_{0})} & = & 1 + H_{0} (t-t_{0}) -
\frac{q_{0}}{2} H_{0}^{2} (t-t_{0})^{2} + \frac{j_{0}}{3!}
H_{0}^{3} (t-t_{0})^{3}  \nonumber \\
& + & \frac{s_{0}}{4!} H_{0}^{4} (t-t_{0})^{4} + \frac{l_{0}}{5!} H_{0}^{5} (t-t_{0})^{5} +\emph{O}[(t-t_{0})^{6}]
\end{eqnarray}
where  the \textit{Hubble} ($H$), \textit{deceleration} ($q$), \textit{jerk} ($j$), \textit{snap} ($s$) and \textit{lerk} parameters are defined, respectively, as
\begin{equation}\label{eq:cosmopar}
H(t) = \frac{\dot{a}}{a}, \quad
q(t) = - \frac{1}{H^{2}} \frac{\ddot{a}}{a}, \quad
j(t) = \frac{1}{H^{3}} \frac{\dddot{a}}{a}, \nonumber
\end{equation}
\begin{equation}\label{eq:cosmopar}
s(t) = \frac{1}{H^{4}} \frac{\ddddot{a}}{a}, \quad
l(t) = \frac{1}{H^{5}} \frac{1}{a}\frac{\mathrm{d}^5 a}{\mathrm{d}t^5} \; ,
\end{equation}
and a dot stands for time derivative.

In Ref.~\cite{SalzanoLazkoz,Xia:2011iv}, the authors developed a comprehensive cosmographic analysis, paying particular attention to two important issues concerning this approach, namely, the convergence of the cosmographic series and the relation between the truncation order of the series and the redshift extent of observational data. The former issue is an increasing source of concern since cosmological observations at $z \gtrsim 1$ are clearly beyond the convergence radius of Eq.~(\ref{eq:aseries}).  A possible way to circumvent this problem was proposed in Ref.~\cite{Cattoen07b}. It consists in parameterizing cosmological distances with a new variable
\begin{equation}
\zeta= \frac{z}{(1+z)}\; ,
\end{equation}
so that the $z$-redshift interval $[0,\infty]$ converts in the $\zeta$-redshift interval $[0,1]$, which encompasses the entire range of observations.

The latter issue arises because a low-order truncation of Eq.~(\ref{eq:aseries}) can lead to wrong estimates of the cosmographic parameters. Naturally, the higher the order considered the more accurate approximation is obtained. However, this can be done only at the expenses of increasing the number of parameters so that a very important aspect involving cosmographic analyses is to determine the order of the expansion which maximizes the statistical significance of the fit for a given data set (for some recent cosmographic analyses, see also \cite{cosmography} and references therein).

\begin{figure*}[htbp]
\centering
\includegraphics[width=15cm]{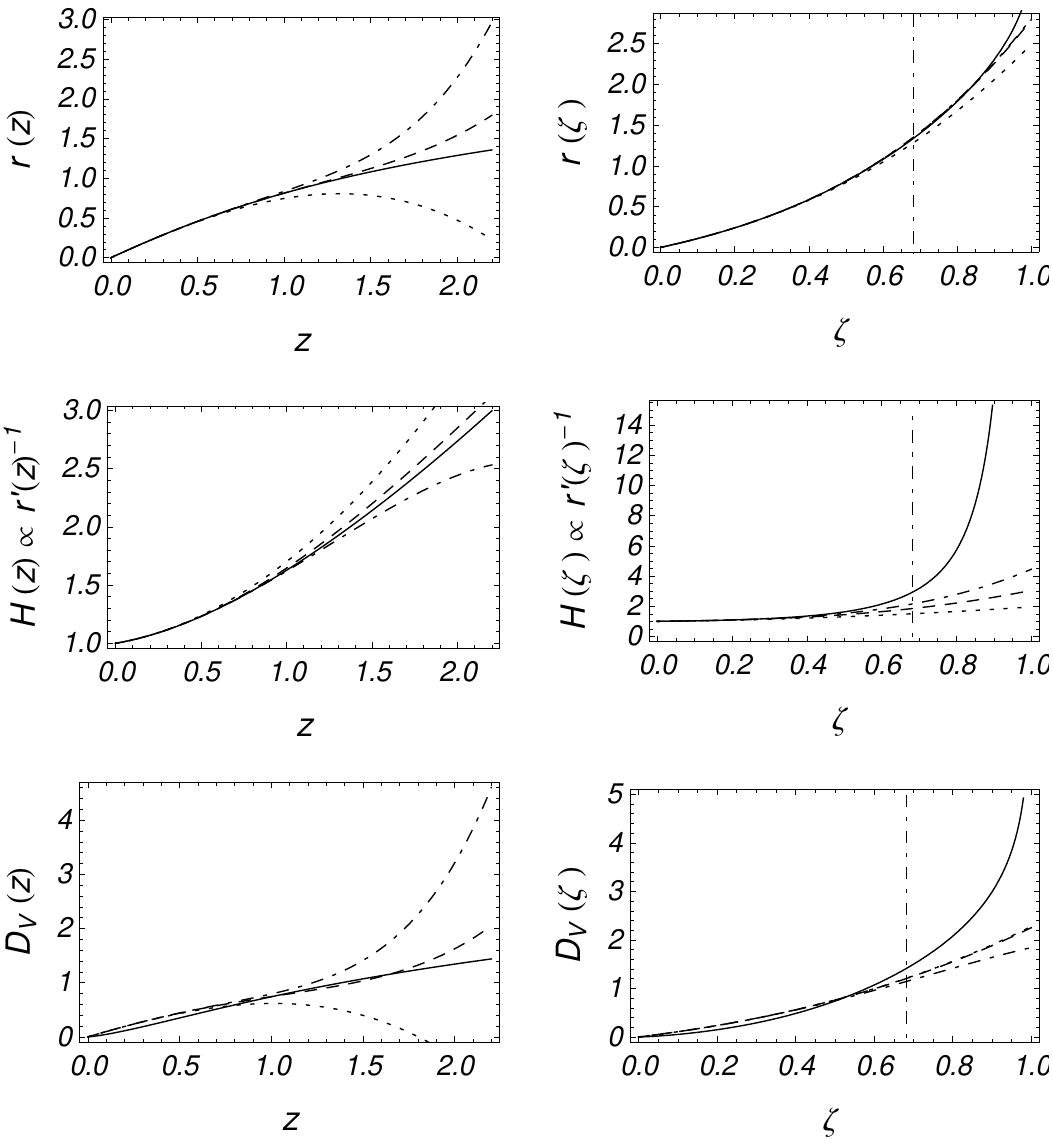}
\caption{Showdown of the exact expressions for BAO quantities with cosmographic series for the fiducial cosmological model used with $\Omega_{m}=0.259$ and $w=-1.12$. \textit{Top:} For $z$-redshift; \textit{Bottom:} For $\zeta$-redshift. \textit{Top column:} dimensionless transverse BAO mode $y$ ($\propto r$); \textit{Center column:} dimensionless radial BAO mode $y'$($\propto r'$); \textit{Bottom column:} dimensionless dilation scale $D_{V}$. The solid line is the exact analytical expressions for the related quantities whereas dotted, dashed and dotdashed lines correspond, respectively, to two, three and four cosmographic parameters series. The vertical dotdashed line is the maximum redshift achievable with \textit{Euclid}, i.e. $z=2.15$ or equally $\zeta =0.68$.}
\label{fig:cosmo_prel}
\end{figure*}

\section{BAO}


Observations of baryonic acoustic oscillations in the clustering of galaxies provide an attractive method for setting cosmological constraints~\cite{bao,wiggle}. In this regard, upcoming BAO measurements may provide  larger and more precise data sets and, consequently, narrower constraints on dark energy by decoupling the BAO transverse $y$ and radial $y'$ modes, defined as \cite{Blake2005}
\begin{equation}\label{eq:yyprima}
y(z) \equiv \frac{r(z)}{r_{s}(z_{r})} \qquad \mathrm{and} \qquad y'(z) \equiv
\frac{r'(z)}{r_{s}(z_{r})} \; ,
\end{equation}
where
\begin{itemize}
  \item $r(z)$ is the comoving distance to the redshift slice $z$;
  \item $r'(z) \equiv \mathrm{d}r(z) / \mathrm{d}z = c/(H(z))$ is the derivative of $r(z)$;
  \item $r_{s}(z_{r})$ is the sound horizon at recombination:
        \begin{equation}
        r_{s}(z_{r}) = \frac{1}{H_{0}} \int_{z_{r}}^{\infty} \frac{c_{s}(z)}{E(z)} \mathrm{d}z\; ,
        \end{equation}
        with $c$ the light velocity, $c_{s}$ the sound speed and $z_{r}$ the recombination
        epoch redshift.
\end{itemize}
Clearly, $y'(z)^{-1} \propto H(z)$ and $y(z) \propto r(z)$; so that we can write these quantities (i.e. $H(z)$ and $r(z)$) as cosmographic series, namely, series in the $z-$  or $\zeta-$redshift (see Appendix A).

Most of the current BAO data are obtained in different way. The position of the BAO approximately constrains the ratio as: $d_{z} \equiv r_{s}(z_{d})/D_{V}(z)$, where $r_{s}(z_{d})$ is the comoving sound horizon at the baryon drag epoch. Since current BAO data are not accurate enough for extracting the information of the angular diameter distance $D_{A}(z)$ and $H(z)$ separately, one can only determine an effective volume distance, defined as~\cite{Eisenstein}
\begin{equation}
D_{V}(z,\boldsymbol{\theta}) = \left[ (1+z)^2 D^{2}_{A} \frac{c\, z}{H(z,\boldsymbol{\theta})} \right]^{1/3}\; ,
\end{equation}
where $D_{A}$ is the angular diameter distance 
and $\boldsymbol{\theta}$ stands for the set of cosmological and/or cosmographic parameters. The cosmographic expressions for the above quantity are provided in the Appendix A.

\subsection{Preliminary analysis}

We performed a preliminary qualitative analysis in order to find out which is the optimal order of cosmographic series one should adopt when working with BAO data lying in a given redshift ($z$ and $\zeta$) interval. To calculate the exact cosmological analytical expressions of all the quantities defined above, we assumed a fiducial model given by the WMAP7-year analysis \cite{wcdm}, i.e., a flat quiessence model \citep{Sahni03} characterized by $\Omega_{m} = 0.259^{+0.099}_{-0.095}$ and $w=-1.12^{+0.42}_{-0.43}$. As explained in Sec. II-B, such a model will also be used for creating our mock BAO data sample.

Then we compare such exact expressions with the corresponding cosmographic series up to the jerk (two cosmographic parameters), snap (three cosmographic parameters) and lerk (four cosmographic parameters) order. From the cosmological fiducial model we used to build up our data, we can easily find out what are the cosmographic parameter values expected for this model (the procedure is explained in Appendix B):
\begin{equation}\label{eq:derived_par}
q_0 = -0.755^{+0.495}_{-0.504},\quad j_0 = 1.448^{+1.738}_{-1.779}, \quad \nonumber
\end{equation}
\begin{equation}
s_0 = 0.730^{+4.143}_{-4.235}, \quad l_0 = 4.152^{+8.478}_{-8.678} .
\end{equation}
The errors are derived with propagation of the errors on the cosmological parameters, $\Omega_{m}$ and $w_{0}$, and clearly show what is the main difficulty in using Cosmography: even very well constrained cosmological parameters can correspond to large uncertainties in the cosmographic ones, being larger for higher order terms.

The results are shown in Figure~\ref{fig:cosmo_prel}. Considering the redshift range of a \textit{Euclid}-like BAO data, two main conclusions can be addressed: $(i)$ it is absolutely necessary to conduct the analysis in the $\zeta$-redshift, as the $z$-redshift fails in the entire redshift range we want to explore; and $(ii)$ {In principle}, it is feasible and useful to extend the cosmographic analysis up to the snap and lerk order, namely with three and four cosmographic parameters.

The transverse BAO mode $y$ can be very well described with the third and fourth order expressions being practically equivalent in the redshift range considered. On the other hand, some aspects must be considered when dealing with the radial BAO mode $y'$. We note that its use can be quite problematic, as the cosmographic series start to deviate in a consistent way from the exact expression at $\zeta \approx 0.4$ ($z \approx 0.67$), just where the sensibility of the {\textit{Euclid}} survey makes the errors on the observed quantity more stringent. Even the actually measured quantity $D_{V}$ shows a clear difficulty in matching cosmographic series with exact analytical expression. All these features could translate in a non-optimal use of Cosmography for such cases.

\subsection{Mock data and statistical analysis}

In order to perform our cosmographic analysis with BAO we generated our mock data based on the fiducial model described earlier. As done in \cite{EscamillaRivera:2011qb} we used the Initiative for Cosmology (\textit{iCosmo}) tool \cite{icosmo}. According to specifications of \textit{iCosmo}, $y$ and $y'$ are calculated by using the fitting formulae of \cite{Blake:2005jd}, with the sound horizon $r_{s}(z_{r})$ derived from \cite{baomock}, once one gives the detail of the survey and a fiducial model. We have simulated the {\textit{Euclid}}-like survey \cite{euclid}, which is expected to cover $15000$ sq. deg. and observe $6.1\times10^7$ galaxies in the redshift range $0.6<z<2.1$. We have used the distribution of galaxies in \cite{Geach}.
We remind that, in order to study the simulated observable quantity $y$, from its definition we also need to fix the value of $H_0 = 74.2$ km s$^{-1}$ Mpc$^{-1}$ \cite{Riess09}, of the sound horizon, $r_{s} = 143.738$ Mpc (derived from relation in \cite{
baomock} using our fiducial cosmological parameters) and the speed of light $c = 299792$ km s$^{-1}$. The choice of $H_{0}$ is completely non influential for results, and the same holds true for the sound horizon $r_{s}(z_{r})$, which appears as the same factor both in mock data and the theoretical quantity to be fitted. Furthermore, the procedure we have followed to create our mock data (described below) automatically implements possible variations of $r_{s}$ in the final total error on $y$ and $y'$. In order to make our mock data not too close to the fiducial model and to give them a more realistic dispersion, we have chosen the following algorithm: we realized $N = 200$ simulations, randomly extracting the fiducial model parameters in the range depicted by their errors; then, for any redshift value derived from the chosen galaxy distribution, we extract the mean and the dispersion of the $y$ and $y'$ distributions, and we use them to characterize a Gaussian probability function. From this last one, we
randomly pick up the values reported in the Table~\ref{tab:dataeuclid}.

With these data, we can define the $\chi^{2}$ function as:
\begin{equation}
\chi^{2}(\theta) = \sum_{i=1}^{\mathcal{N}_{BAO}} \frac{(\mathcal{F}(\zeta_{i}, \boldsymbol{\theta}) - \mathcal{F}_{obs}(\zeta_{i}))^{2}}{\sigma_{i}^{2}},
\end{equation}
where $\boldsymbol{\theta}$ is the cosmographic parameters vector, and $\mathcal{F}$ will be in this case $y(\zeta,\boldsymbol{\theta})$, as we have specified in the previous section. To determine the best fit cosmographic parameters we use a Markov Chain Monte Carlo (MCMC) method, following \cite{Dunkley:2004sv} to test the convergence of each chain. We have tried to be as general as possible at the moment of giving a physically viable prior, having finished with the only condition that $y(\zeta) >0$ all over the $\zeta$-redshift range $[0;1]$, which is a very general and natural choice. Finally, we performed our analysis in three different redshift ranges: $0.1 \leq z \leq 2.15$, i.e.,  assuming that a survey with a {\textit{Euclid}}-like sensibility can cover all this range; $0.6 \leq z \leq 2.15$, which is the effective observable range of the {\textit{Euclid}} mission; and $0.1 \leq z \leq 0.75$, which is the range actually covered by current BAO observations \cite{wiggle}.

In order to statistically compare our results, we calculate two quantities. One is the Figure of Merit (FoM), which is generally defined as the $N$-dimensional volume enclosed by the confidence level contours of the parameters $\boldsymbol{\theta}$ and written as: $FoM_{\boldsymbol{\theta}} = 1/\sqrt{\mathrm{det} \; Cov(\boldsymbol{\theta})} $ \citep{Wang08}, with $Cov(\boldsymbol{\theta})$ the covariance matrix of the considered theoretical parameters. We calculate the FoM for $(q_{0},j_{0})$, given that these are the only well-constrained cosmographic parameters.

The other quantity is the Bayesian evidence, defined as the probability of the data $D$ given the model $M$ with a set of parameters $\boldsymbol{\theta}$, $\mathcal{E}(M) = \int \mathrm{d}\boldsymbol{\theta} L(D|\boldsymbol{\theta},M)\pi(\boldsymbol{\theta}|M)$:
$\pi(\boldsymbol{\theta}|M) $ is the prior on the set of parameters, normalized to unity, and $L(D|\boldsymbol{\theta},M)$ is the likelihood function. We impose flat priors on the estimated cosmographical parameters over sufficiently wide ranges so that further increasing these ranges has no impact on the results. The evidence is estimated using the algorithm given in Ref.~\citep{evidence}; to reduce the statistical noise we run the algorithm several  times to obtain a distribution of $\sim 100$ values from which we extract the best value of the evidence as the mean of such a distribution. Then, we calculate the Bayes Factor, which is defined as the ratio of evidence of two models, $M_{i}$ and $M_{j}$, $\mathcal{B}_{ij} = \mathcal{E}_{i}/\mathcal{E}_{j}$. If $\mathcal{B}_{ij} > 1$, model $M_i$ is preferred over $M_j$, given the data. In our case, the reference model with the highest evidence is the two dimensional cosmographic series, for each one of redshift ranges we considered.

As in the frequentist approach, where one compares the $\chi^2$ of given models and concludes how significantly a model fits the data, in the Bayesian evidence framework one can assign significance to the difference between models by using the Jeffreys Scale: if $\ln \mathcal{B}_{ij} < 1$ there is not significant preference for the model with the highest evidence; if $1<\ln \mathcal{B}_{ij} < 2.5$ the preference is substantial; if $2.5<\ln \mathcal{B}_{ij} < 5$ it is strong; if $\ln \mathcal{B}_{ij} > 5$ it is decisive.

\section{Results and discussion}

\begin{figure*}[htbp]
\centering
\includegraphics[width=8.2cm]{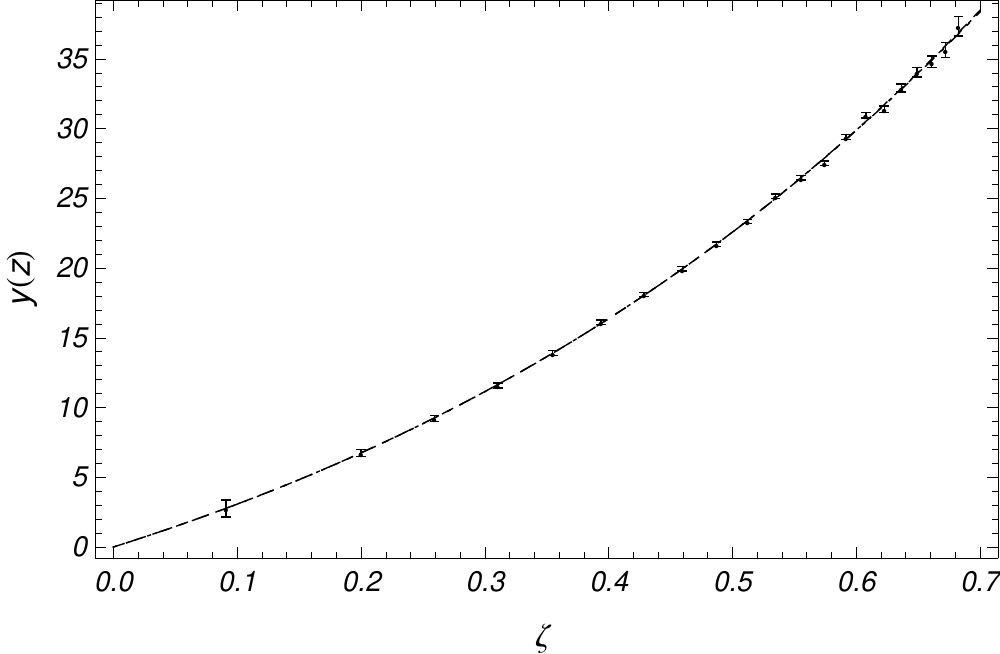}\\
~~~\\
\includegraphics[width=8.2cm]{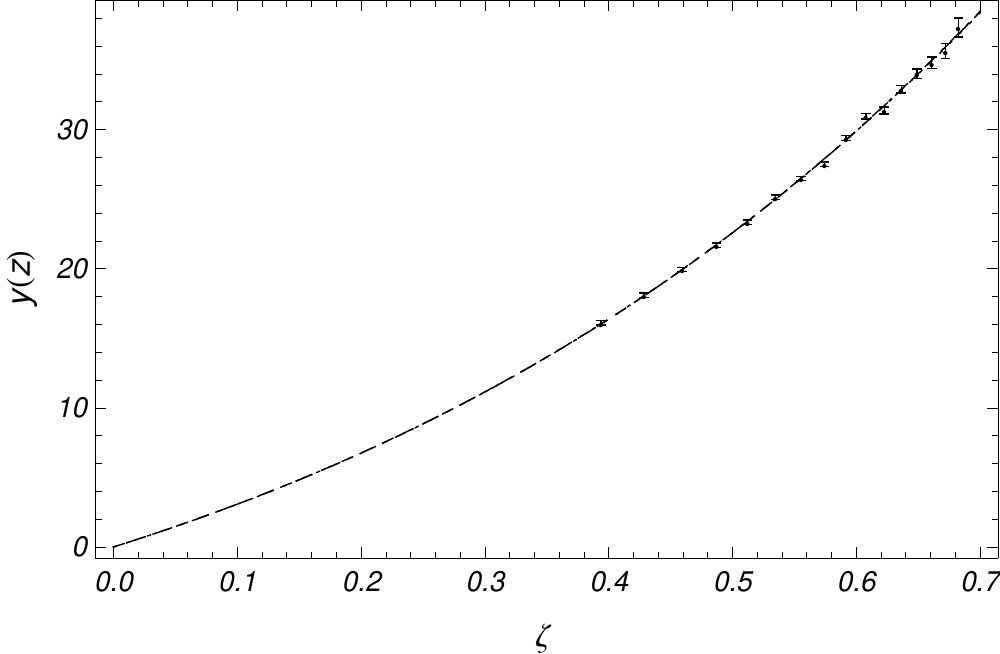}\\
~~~\\
\includegraphics[width=8.2cm]{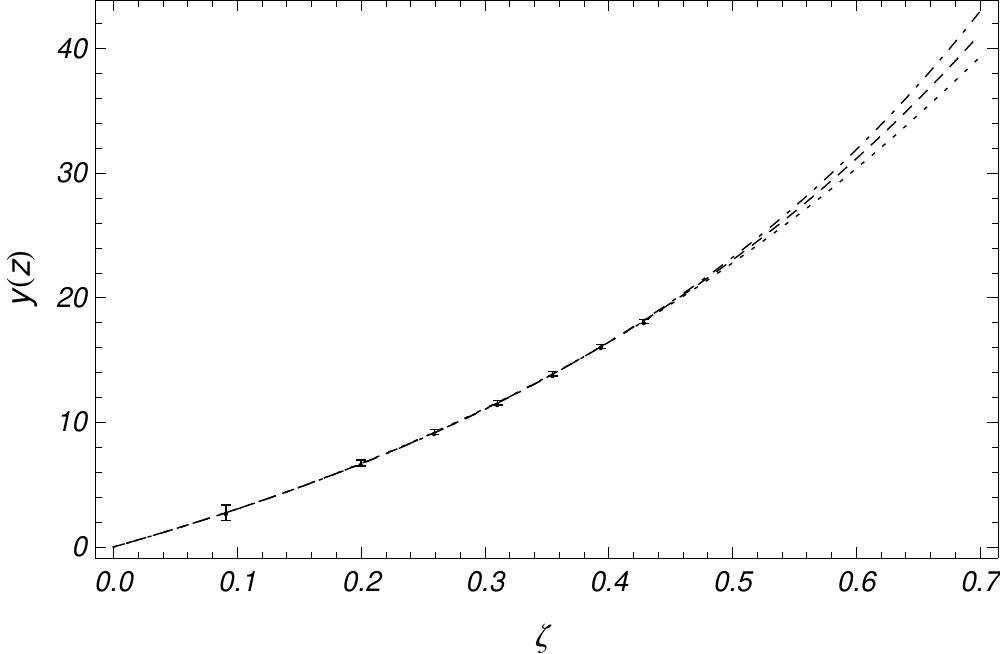}
\caption{Showdown of $y$ from cosmographic best-fits versus \textit{Euclid} simulated data. Dashed line: two parameters cosmographic series; dotted line: three parameters cosmographic series; dot-dashed line: four cosmographic parameters series. \textit{Top panel:} results for the entire redshift range, $[0.1;2.15]$. \textit{Center panel:} results considering future \textit{Euclid} redshift range, $[0.6;2.15]$; \textit{Bottom panel:} results for the present observational BAO redshift range, $[0.1;0.75]$.}
\label{fig:cosmo_end}
\end{figure*}

How can we quantify if the use of BAO data as a cosmographic tool is giving us reasonably results? The easiest way is to compare the derived cosmographic best fit values shown in Table~\ref{tab:bestfits} with the expected ones in Eq.~(\ref{eq:derived_par})  -- see also Fig.~\ref{fig:cosmo_end}. We can see that, the BAO-reconstructed cosmographic parameter values are perfectly compatible with the {expected} fiducial values, with differences arising mainly because of the realistic dispersion of data that we have added to our mock sample. It is also clear that, even when such high-redshift BAO data are implemented, we are not able to constrain more than two cosmographic parameters, i.e., deceleration and jerk, while snap and lerk remain out of our possibilities. This is clearly showed by the fact that snap and lerk best-fit estimations are very centered in zero and with large tails in the probability distribution, and differ from the minimum $\chi^2$ values; a typical feature of MCMC which means that
these parameters cannot be constrained. The opposite happens for deceleration and jerk.

The FoM values clearly confirm that as long as we add cosmographic parameters, the parameter space volume increases and the constraining power decreases through degeneracies between the parameters. The Bayes factor also does not show no striking evidence in favor of the two dimensional case.

Notwithstanding, the use of high order expressions is fundamental in order to soften possible biases in the estimation of deceleration and jerk. Indeed, we can see that estimates from a two-parameter cosmographic series are very different from the expected fiducial values. On the other hand, extending the series up to higher order makes us able to get just the expected values (inside the $1\sigma$ confidence levels) at least for what it concerns deceleration and jerk, still laking enough observational accuracy in order to assess a statistically valid estimation for the other higher-order cosmographic parameters. 

{\renewcommand{\tabcolsep}{3.mm}
{\renewcommand{\arraystretch}{1.25}
\begin{table}
\begin{minipage}{\textwidth}
\caption{\textit{\textit{Euclid}} Mock data distribution for BAO}\label{tab:dataeuclid}
\centering
\resizebox*{0.5\textwidth}{!}{
\begin{tabular}{ccccc}
\hline
\hline
{Redshift}  &  {$y$}  & {$\sigma_{y}$} & {$y'$} & {$\sigma_{y'}$}\\
\hline
\hline
$0.1$   & $2.758$   & $0.616$ & $27.153$ & $3.676$ \\
$0.25$  & $6.742$   & $0.250$ & $25.449$ & $1.477$ \\
$0.35$  & $9.214$   & $0.200$ & $24.877$ & $0.892$ \\
$0.45$  & $11.578$   & $0.180$ & $23.147$ & $0.617$ \\
$0.55$  & $13.904$   & $0.169$ & $22.347$ & $0.462$ \\
$0.65$  & $16.107$   & $0.162$ & $20.915$ & $0.364$ \\
$0.75$  & $18.105$   & $0.158$ & $19.681$ & $0.299$ \\
$0.85$  & $19.938$   & $0.156$ & $18.496$ & $0.252$ \\
$0.95$  & $21.699$   & $0.156$ & $17.347$ & $0.218$ \\
$1.05$  & $23.341$    & $0.157$ & $16.583$ & $0.191$ \\
$1.15$  & $25.138$    & $0.158$ & $15.434$ & $0.171$ \\
$1.25$  & $26.481$    & $0.160$ & $14.744$ & $0.154$ \\
$1.35$  & $27.515$    & $0.169$ & $13.815$ & $0.147$ \\
$1.45$  & $29.381$     & $0.185$ & $13.207$ & $0.145$ \\
$1.55$  & $30.963$     & $0.209$ & $12.481$ & $0.149$ \\
$1.65$  & $31.371$     & $0.240$ & $11.904$ & $0.156$ \\
$1.75$  & $32.904$     & $0.281$ & $11.217$ & $0.168$ \\
$1.85$  & $34.028$     & $0.338$ & $10.899$ & $0.186$ \\
$1.95$  & $34.790$     & $0.417$ & $10.294$ & $0.212$ \\
$2.05$  & $35.645$     & $0.529$ & $9.752$ & $0.250$ \\
$2.15$  & $37.341$     & $0.693$ & $9.344$ & $0.303$ \\
\hline
\hline
\end{tabular}}
\end{minipage}
\end{table}}}

{\renewcommand{\tabcolsep}{1mm}
{\renewcommand{\arraystretch}{2.5}
\begin{table*}
\begin{minipage}{\textwidth}
 \caption{Best fits cosmographic parameters.  \textit{Column 1:} series order: $2D - $ two cosmographic parameters; $3D - $ three cosmographic parameters; $4D - $ four cosmographic parameters. \textit{Column 2:} redshift range. \textit{Column 3-4-5:} best fit estimations (corresponding to the median of MCMC-derived probability distributions). \textit{Column 6-7-8:} values of cosmographic parameters corresponding to the minimum $\chi^2$; if they differ from median-derived ones, the parameter is not constrained. \textit{Column 9:} figure of merit for deceleration and jerk ($FoM_{q_{0}j_{0}} = 1/\sqrt{\mathrm{det} \; Cov_{q_{0}j_{0}}} $). \textit{Column 10:} Bayes factor with respect to the $2-D$ case for each redshift range case.}\label{tab:bestfits}
\resizebox*{\textwidth}{!}{
 \begin{tabular}{c|c|cccc|cccc|cc}
  \hline
  \hline
    $z-$range  &          &  $q_{0}$ & $j_{0}$ & $s_{0}$ & $l_{0}$ & $q_{0}^{min}$ & $j_{0}^{min}$ & $s_{0}^{min}$ & $l_{0}^{min}$ & FoM$_{q_{0}j_{0}}$ & $\ln \mathcal{B}_{ij}$ \\
  \hline
  \hline
               & $y-2D$   & $-0.696^{+0.086}_{-0.084}$ & $0.462^{+0.987}_{-0.966}$ & $-$ & $-$ & $-0.697$ & $0.467$ & $-$ & $-$ & $525.57$ & $0$ \\
  $0.1 - 2.15$ & $y-3D$   & $-0.764^{+0.051}_{-0.046}$ & $1.774^{+0.296}_{-0.321}$ & $\mathit{0.006^{+0.233}_{-0.149}}$ & $-$ & $-0.758$ & $1.595$ & $\mathit{-2.550}$ & $-$ & $489.19$ & $0.21$ \\
               & $y-4D$   & $-0.748^{+0.035}_{-0.039}$ & $1.588^{+0.210}_{-0.200}$ & $\mathit{0.004^{+0.544}_{-0.346}}$ & $\mathit{-0.050^{+0.183}_{-0.692}}$ & $-0.774$ & $1.939$ & $\mathit{2.232}$ & $\mathit{-0.127}$ & $98.84$ & $0.23$\\
  \hline
  \hline
                & $y-2D$   & $-0.711^{+0.101}_{-0.101}$ & $0.627^{+1.194}_{-1.134}$ & $-$ & $-$ & $-0.710$ & $0.614$ & $-$ & $-$ & $408.53$ & $0$\\
  $0.65 - 2.15$ & $y-3D$   & $-0.769^{+0.057}_{-0.052}$ & $1.807^{+0.326}_{-0.365}$ & $\mathit{0.005^{+0.179}_{-0.207}}$ & $-$ & $-0.785$ & $1.954$ & $\mathit{0.824}$& $-$ & $351.68$ & $0.15$\\
                & $y-4D$   & $-0.740^{+0.092}_{-0.047}$ & $1.535^{+0.239}_{-1.322}$ & $\mathit{-0.044^{+0.352}_{-7.710}}$ & $\mathit{-0.010^{+0.108}_{-0.193}}$ & $-0.758$ & $1.711$ & $\mathit{1.022}$ & $\mathit{-0.113}$ & $14.89$ & $0.20$\\
\hline
\hline
               & $y-2D$   & $-0.565^{+0.324}_{-0.322}$ & $-1.434^{+4.502}_{-3.913}$ & $-$ & $-$ & $-0.565$ & $-1.438$ &$-$ & $-$ & $33.23$ & $0$\\
  $0.1 - 0.75$ & $y-3D$   & $-0.565^{+0.033}_{-0.040}$ & $\mathit{0.016^{+0.293}_{-0.177}}$ & $\mathit{0.009^{+0.327}_{-0.220}}$ & $-$ & $-0.634$ & $\mathit{0.030}$ & $\mathit{-11.36}$ & $-$ & $-$ & $0.06$\\
               & $y-4D$   & $-0.536^{+0.032}_{-0.051}$ & $\mathit{0.020^{+0.310}_{-0.149}}$ & $\mathit{-0.0004^{+0.2084}_{-0.2368}}$ & $\mathit{-0.004^{+0.210}_{-0.232}}$ & $-0.617$ & $\mathit{-0.018}$ & $\mathit{-7.759}$ & $\mathit{0.065}$ & $-$ & $0.27$ \\
\hline
\hline
\end{tabular}}
\end{minipage}
\end{table*}}}

\section{Final Remarks}

In this paper, we have discussed BAO cosmography using simulated data from a \textit{Euclid}-like survey~\cite{euclid}. We have provided general expressions for the transversal and radial BAO modes $y$ and $y'$ and discussed the feasibility of extending cosmographic series up to higher orders. As mentioned earlier, finding the optimal order of these expansions is very important to reliably reconstruct the cosmic scale factor in a model-independent way.

We have shown that  high-redshift BAO data can place tight constraints on the deceleration and jerk parameters (see Table \ref{tab:bestfits}), which allows a ({\it{i}}) model-independent check of the cosmic acceleration ($q_0 <0$) and ({\it{ii}}) a discrimination between the standard $\Lambda$CDM model ($j_0 = 1$) and alternative mechanisms of cosmic acceleration. As discussed, the next generation of redshift surveys will provide data sets with sufficient accuracy to map the history of cosmic expansion in a model-independent way. In this regard, the forecasts
presented here clearly show that BAO cosmography can be useful.


\acknowledgments

RL, VS and IS are supported by the Spanish Ministry of Economy and Competitiveness through research projects FIS2010-15492 and Consolider EPI CSD2010-00064, and also by the Basque Government through research project   IT592-13, and by the University of the Basque Country UPV/EHU under program UFI 11/55. C. Escamilla-Rivera is supported by Fundaci\'on Pablo Garc\'ia-FUNDEC, M\'exico. Jailson Alcaniz is supported by CNPq (305857/2010-0 and 485669/2011-0) and FAPERJ (E-26/103.239/2011).


\section{Appendix A}
\label{sec:supportingA}

We can express the BAO modes as cosmographic series, i.e., series in the $\zeta-$redshift depending on the cosmographic parameters evaluated at $\zeta=0$. We have, for the $\zeta$-redshift, the dimensionless quantities (out of a factor $c/H_{0}$):
\begin{eqnarray}\label{H_zetaeq}
&&H(\zeta,q_0 ,j_0,s_0,l_{0})\propto \mathcal{H}_{0}^{\zeta} + \mathcal{H}_{1}^{\zeta} \zeta + \mathcal{H}_{2}^{\zeta} \zeta^2 + \mathcal{H}_{3}^{\zeta} \zeta^3 + \mathcal{H}_{4}^{\zeta} \zeta^4 \\
&&\mathcal{H}_{0}^{\zeta}= 1 \nonumber \\
&&\mathcal{H}_{1}^{\zeta}= 1+q_0 \nonumber \\
&&\mathcal{H}_{2}^{\zeta}= \frac{1}{2}(2+2q_0 - q_0^2+j_0) \nonumber \\
&&\mathcal{H}_{3}^{\zeta}= \frac{1}{6}(6 + 3 q_0^3 - 3 q_0^2 + 6 q_0 + 3 j_0  - 4 j_0 q_0 - s_0 )\nonumber \\
&&\mathcal{H}_{4}^{\zeta}= \frac{1}{24}(24 - 15 q_0^4 + 12 q_0^3 - 12 q_0^2 + 24 q_0 - 4 j_0^2 + 12 j_0 - 16 j_0 q_0 + 25 j_0 q_0^2 - 4 s_0 + 7 q_0 s_0 \nonumber \\
&& + l_0); \nonumber
\end{eqnarray}
\begin{eqnarray}\label{r_zetaeq}
&&r(\zeta,q_0,j_0,s_0,l_{0}) = \mathcal{R}_{1}^{\zeta} \zeta + \mathcal{R}_{2}^{\zeta} \zeta^2 + \mathcal{R}_{3}^{\zeta} \zeta^3 + \mathcal{R}_{4}^{\zeta} \zeta^4 + \mathcal{R}_{5}^{\zeta} \zeta^5 \\
&&\mathcal{R}_{1}^{\zeta}= 1 \nonumber \\
&&\mathcal{R}_{2}^{\zeta}= \frac{1}{2}(1-q_0) \nonumber \\
&&\mathcal{R}_{3}^{\zeta}= \frac{1}{6} (2 -2q_0 +3 q_0^2-j_0)\nonumber \\
&&\mathcal{R}_{4}^{\zeta}= \frac{1}{24}(6 -6 q_0 +9 q_0^2 -15 q_0^3 -3 j_0 + 10 q_0 j_0 + s_0) \nonumber \\
&&\mathcal{R}_{5}^{\zeta}= \frac{1}{120} (24 -24 q_{0} + 36 q_{0}^2 - 60 q_{0}^3 + 105 q_{0}^4 - 12 j_{0} + 10 j_{0}^2 +40 q_{0}j_{0} -105 q_{0}^2 j_{0}+ 4 s_{0} \nonumber \\
&& -15 q_{0}s_{0} - l_{0});\nonumber
\end{eqnarray}
and
\begin{eqnarray}\label{DV_zetaeq}
&&D_{V}(\zeta,q_0,j_0,s_0,l_{0}) = \mathcal{D}_{1}^{\zeta} \zeta + \mathcal{D}_{2}^{\zeta} \zeta^2 + \mathcal{D}_{3}^{\zeta} \zeta^3 + \mathcal{D}_{4}^{\zeta} \zeta^4 + \mathcal{D}_{5}^{\zeta} \zeta^5 \\
&&\mathcal{D}_{1}^{\zeta}= 1 \nonumber \\
&&\mathcal{D}_{2}^{\zeta}= \frac{1}{3}(1-2q_0) \nonumber \\
&&\mathcal{D}_{3}^{\zeta}= \frac{1}{36} (7 -10q_0 +29 q_0^2-10j_0)\nonumber \\
&&\mathcal{D}_{4}^{\zeta}= \frac{1}{324}(44 -57 q_{0} +117 q_{0}^2 - 376 q_{0}^3 - 39 j_{0} + 258 q_0 j_0 + 27 s_0) \nonumber \\
&&\mathcal{D}_{5}^{\zeta}= \frac{1}{19440} (2017 -2492 q_{0} + 4638 q_{0}^2 - 10460 q_{0}^3 + 35395 q_{0}^4 - 1536 j_{0} + 3540 j_{0}^2 + 6990 q_{0}j_{0}\nonumber \\
&& -36300 q_{0}^2 j_{0} +702 s_{0} - 5400 q_{0}s_{0}); \nonumber
\end{eqnarray}
The $z-$redshift expressions can be easily obtained by performing the variable change $\zeta \rightarrow z$ and making the series expansion.

\section{Appendix B}
\label{sec:supportingB}

The fiducial model used in the work is a quiessence cosmological model where the dimensionless Hubble function reads:
\begin{equation}
E^2(z) = \frac{H^{2}(z)}{H^{2}_{0}} = \Omega_m (1 + z)^3 + \Omega_x (1 + z)^{3(1 + w_0)}, \label{eq: ecpl}
\end{equation}
In order to determine the cosmographic parameters for such a model, we avoid integrating $H(z)$ to get $a(t)$ by noting that $d/dt = -(1 + z) H(z) d/dz$. We can use such a relation to evaluate $(dH/dt, d^2H/dt^2, d^3H/dt^3)$ and convert the Eqs. (\ref{eq:cosmopar}) to redshift derivatives:
\begin{equation}
q(z) = \frac{(1+z)}{2 H^2 (z)}\frac{dH^2 (z)}{dz} -1,
\end{equation}
\begin{equation}
j(z) = \frac{(1+z)^2}{2 H^2 (z)}\frac{d^2 H^2 (z)}{dz^2} -\frac{(1+z)}{2 H^2 (z)}\frac{dH^2 (z)}{dz} +1\,.
\end{equation}
We then solve above equations and evaluate them at $z= 0$. For the snap and the lerk parameters the expressions are much longer and involve the third and fourth order derivatives of $H^2(z)$. After some algebra, and using the cosmological parameters $\Omega_{m}$ and $w_{0}$ that characterize the chosen fiducial model, we have the set of values described in the paper.



\vfill

\begin{thebibliography}{99}

\bibitem{SNfirst}
Riess et al., ApJ 116 (1998) 1009;
S. Perlmutter, et al., ApJ 517 (1999) 565.

\bibitem{SNsecond} V. Sahni and A. A. Starobinsky, Int. J. Mod. Phys. D9,
373 (2000); P. J. E. Peebles and B. Ratra, Rev. Mod. Phys. 75, 559 (2003); T. Padmanabhan, Phys. Rept.
380, 235 (2003); J.~S.~Alcaniz, Braz.\ J.\ Phys.\  {\bf 36}, 1109 (2006); E.~J.~Copeland, M.~Sami, S.~Tsujikawa, Int. J. Mod. Phys. D15 (2006) 1753; M. Bartelmann, Rev. Mod. Phys. 82, 331 (2010).

\bibitem{euclid}  R.~Laureijs {\it et al.}  [EUCLID Collaboration],
  arXiv:1110.3193 [astro-ph.CO];
A. Refregier, et al., arXiv:1001.0061;
J.P. Beaulieu, et al., arXiv:1001.3349;
M. Martinelli, et al., Phys. Rev. D 83 (2011) 023012.

\bibitem{Moles:2009db}
M.~Moles, et al., PASP 122 (2010) 363; {http://j-pas.org}.

\bibitem{Eisenstein} D.~J.~Eisenstein {\it et al.}  [SDSS Collaboration],
  Astrophys.\ J.\  {\bf 633} (2005) 560.

\bibitem{bao}
Y. Wang, P. Mukherjee, Phys. Rev. D 76 (2007) 103533;
P.S. Corasaniti, A. Melchiorri, Phys. Rev. D 77 (2008) 103507;
Percival, W.J., et al., MNRAS 401 (2010) 2148;
Beutler, F., et al., MNRAS 416 (2011) 3017;
H.~-~J.~Seo, et al., ApJ 761 (2012) 13.

\bibitem{wiggle}
C.~Blake, E.~A.~Kazin, F.~Beutler, et al., MNRAS 418 (2011) 1707;


\bibitem{SalzanoLazkoz}
Capozziello, S., Lazkoz, R., Salzano, V., Phys. Rev. D 84 (2011) 124061.

\bibitem{Xia:2011iv}
J.~-Q.~Xia, V.~Vitagliano, S.~Liberati, M.~Viel, Phys. Rev. D 85 (2012) 043520.

\bibitem{Cattoen07b}
Cattoen, C., Visser, M., Classic and Quantum Gravity 24 (2007) 5985;
Cattoen, C., Visser, M., 2007, arXiv:gr-qc/0703122.


\bibitem{cosmography} M.~Visser,
  Gen.\ Rel.\ Grav.\  {\bf 37}, 1541 (2005); C.~Cattoen and M.~Visser,
  Phys.\ Rev.\ D {\bf 78}, 063501 (2008); M.~Visser and C.~Cattoen,
  arXiv:0906.5407 [gr-qc]; E.~Mortsell and C.~Clarkson,
  JCAP {\bf 0901}, 044 (2009); S.~Capozziello and V.~Salzano,
  Adv.\ Astron.\  {\bf 2009}, 217420 (2009); J.~C.~Carvalho and J.~S.~Alcaniz,
  Mon.\ Not.\ Roy.\ Astron.\ Soc.\  {\bf 418}, 1873 (2011); A.~C.~C.~Guimaraes and J.~A.~S.~Lima,  Class.\ Quant.\                Grav.\  {\bf 28}, 125026 (2011); R.~F.~L.~Holanda, J.~S.~Alcaniz and J.~C.~Carvalho,
  JCAP {\bf 1306}, 033 (2013).

\bibitem{Blake2005}
C. Blake, et al., MNRAS 365 (2006) 255

\bibitem{wmap7matrix}
{http://lambda.gsfc.nasa.gov/product/map/current \\
/parameters.cfm}


\bibitem{Riess09}
A.G. Riess, et al., ApJ 699 (2009) 539.

\bibitem{Geach}
J.E. Geach, et al., MNRAS 699 (2009) 539.

\bibitem{EscamillaRivera:2011qb}
C.~Escamilla-Rivera, R.~Lazkoz, V.~Salzano, I.~Sendra, JCAP 09 (2011) 3.

\bibitem{icosmo}
A.~Refregier, A.~Amara, T.~Kitching, A.~Rassat, A$\&$A 528 (2011) 33.

\bibitem{Blake:2005jd}
C.~Blake, et al., MNRAS  365 (2006) 255.

\bibitem{baomock}
W. Hu, N. Sugiyama, ApJ 471 (1996) 542;
G. Efstathiou, J.R. Bond, MNRAS 304 (1999) 75.

\bibitem{wcdm}
{http://lambda.gsfc.nasa.gov/product/map/dr4/params \\ /wcdm\_sz\_lens\_wmap7.cfm}.

\bibitem{Sahni03}
V. Sahni, T.~D. Saini, A.~A. Starobinsky, U. Alama, JETP Lett. 77 (2003) 201.

\bibitem{Dunkley:2004sv}
J.~Dunkley, M.~Bucher, P.~G.~Ferreira, K.~Moodley, C.~Skordis, MNRAS 356 (2005) 925.

\bibitem{Wang08}
Y. Wang, Phys. Rev. D 77 (2008) 123525.

\bibitem{evidence}
P. Mukherjee, D. Parkinson, A.R. Liddle, Astrophys. J. 638 (2006) 51-51.


\end{thebibliography}
\end{document}